\newfam\scrfam
\batchmode\font\tenscr=rsfs10 \errorstopmode
\ifx\tenscr\nullfont
        \message{rsfs script font not available. Replacing with calligraphic.}
        \def\scr{\cal}
\else   \font\twelvescr=rsfs12
        
        \font\sevenscr=rsfs7
        \font\fivescr=rsfs5
        \skewchar\twelvescr='177
        \skewchar\tenscr='177 \skewchar\sevenscr='177 \skewchar\fivescr='177
        \textfont\scrfam=\twelvescr \scriptfont\scrfam=\tenscr
        \scriptscriptfont\scrfam=\sevenscr
        \def\scr{\fam\scrfam}
        \def\cal{\scr}
\fi
\newfam\msbfam
\batchmode\font\twelvemsb=msbm10 scaled\magstep1 \errorstopmode
\ifx\twelvemsb\nullfont\def\Bbb{\bf}

	\message{Blackboard bold not available. Replacing with boldface.}
\else   \catcode`\@=11
        \font\tenmsb=msbm10 \font\sevenmsb=msbm7 
        \textfont\msbfam=\twelvemsb
        \scriptfont\msbfam=\tenmsb \scriptscriptfont\msbfam=\sevenmsb
        \def\Bbb{\relax\expandafter\Bbb@}
        \def\Bbb@#1{{\Bbb@@{#1}}}
        \def\Bbb@@#1{\fam\msbfam\relax#1}
        \catcode`\@=\active

\fi
        \font\eightrm=cmr8              
        
        \font\teni=cmmi10               
        \font\tencp=cmcsc10
        \font\tentt=cmtt10
        \font\twelverm=cmr12
        \font\twelvecp=cmcsc10 scaled\magstep1
        \font\fourteencp=cmcsc10 scaled\magstep2
        \font\fiverm=cmr5

	 at10pt	
	\font\twelvehelvbold=phvb at12pt
	 at14pt
	\font\sixteenhelvbold=phvb at16pt

\font\tenrm=cmr10    \def\xrm{\tenrm}
\font\tenbf=cmbx10   \def\xbf{\tenbf}
\font\tenit=cmti10   \def\xit{\tenit}
\font\tentt=cmtt10   
\font\teni=cmmi10    \def\xold{\teni}
\font\tenib=cmmib10  \def\xbold{\tenib}
\font\twelvei=cmmi12 \def\old{\twelvei}

%

\font\fourteenrm  =cmr10 scaled\magstep2
\font\twelverm    =cmr12
\font\ninerm      =cmr9
\font\sixrm       =cmr6

\font\fourteenbf  =cmbx10 scaled\magstep2
\font\twelvebf    =cmbx12
\font\ninebf      =cmbx9
\font\sixbf       =cmbx6
\font\seventeeni  =cmmi10 scaled\magstep3    \skewchar\seventeeni='177
\font\fourteeni   =cmmi10 scaled\magstep2     \skewchar\fourteeni='177
\font\twelvei     =cmmi12                       \skewchar\twelvei='177
\font\ninei       =cmmi9                          \skewchar\ninei='177
\font\sixi        =cmmi6                           \skewchar\sixi='177
\font\seventeensy =cmsy10 scaled\magstep3    \skewchar\seventeensy='60
\font\fourteensy  =cmsy10 scaled\magstep2     \skewchar\fourteensy='60
\font\twelvesy    =cmsy10 scaled\magstep1       \skewchar\twelvesy='60
\font\ninesy      =cmsy9                          \skewchar\ninesy='60
\font\sixsy       =cmsy6                           \skewchar\sixsy='60

\font\fourteenex  =cmex10 scaled\magstep2
\font\twelveex    =cmex10 scaled\magstep1

\font\fourteensl  =cmsl10 scaled\magstep2
\font\twelvesl    =cmsl12
\font\ninesl      =cmsl9

\font\fourteenit  =cmti10 scaled\magstep2
\font\twelveit    =cmti12
\font\nineit      =cmti9
\font\fourteentt  =cmtt10 scaled\magstep2
\font\twelvett    =cmtt12
\font\fourteencp  =cmcsc10 scaled\magstep2
\font\twelvecp    =cmcsc10 scaled\magstep1
\font\tencp       =cmcsc10
%
%
\catcode`\@=11 

\def\fourteenf@nts{\relax
    \textfont0=\fourteenrm          \scriptfont0=\tenrm
      \scriptscriptfont0=\sevenrm
    \textfont1=\fourteeni           \scriptfont1=\teni
      \scriptscriptfont1=\seveni
    \textfont2=\fourteensy          \scriptfont2=\tensy
      \scriptscriptfont2=\sevensy
    \textfont3=\fourteenex          \scriptfont3=\twelveex
      \scriptscriptfont3=\tenex
    \textfont\itfam=\fourteenit     \scriptfont\itfam=\tenit
    \textfont\slfam=\fourteensl     \scriptfont\slfam=\tensl
    \textfont\bffam=\fourteenbf     \scriptfont\bffam=\tenbf
      \scriptscriptfont\bffam=\sevenbf
    \textfont\ttfam=\fourteentt
    \textfont\cpfam=\fourteencp }
\def\twelvef@nts{\relax
    \textfont0=\twelverm          \scriptfont0=\ninerm
      \scriptscriptfont0=\sixrm
    \textfont1=\twelvei           \scriptfont1=\ninei
      \scriptscriptfont1=\sixi
    \textfont2=\twelvesy           \scriptfont2=\ninesy
      \scriptscriptfont2=\sixsy
    \textfont3=\twelveex          \scriptfont3=\tenex
      \scriptscriptfont3=\tenex
    \textfont\itfam=\twelveit     \scriptfont\itfam=\nineit
    \textfont\slfam=\twelvesl     \scriptfont\slfam=\ninesl
    \textfont\bffam=\twelvebf     \scriptfont\bffam=\ninebf
      \scriptscriptfont\bffam=\sixbf
    \textfont\ttfam=\twelvett
    \textfont\cpfam=\twelvecp }
\def\tenf@nts{\relax
    \textfont0=\tenrm          \scriptfont0=\sevenrm
      \scriptscriptfont0=\fiverm
    \textfont1=\teni           \scriptfont1=\seveni
      \scriptscriptfont1=\fivei
    \textfont2=\tensy          \scriptfont2=\sevensy
      \scriptscriptfont2=\fivesy
    \textfont3=\tenex          \scriptfont3=\tenex
      \scriptscriptfont3=\tenex
    \textfont\itfam=\tenit     \scriptfont\itfam=\seveni  
    \textfont\slfam=\tensl     \scriptfont\slfam=\sevenrm 
    \textfont\bffam=\tenbf     \scriptfont\bffam=\sevenbf
      \scriptscriptfont\bffam=\fivebf
    \textfont\ttfam=\tentt
    \textfont\cpfam=\tencp }
%
%

%
\let\rel@x=\relax
\let\n@expand=\relax
\def\pr@tect{\let\n@expand=\noexpand}
\let\protect=\pr@tect
\let\gl@bal=\global
%
%
%
\newfam\cpfam
\newdimen\b@gheight             \b@gheight=12pt
\newcount\f@ntkey               \f@ntkey=0
\def\f@m{\afterassignment\samef@nt\f@ntkey=}
\def\samef@nt{\fam=\f@ntkey \the\textfont\f@ntkey\rel@x}
\def\setstr@t{\setbox\strutbox=\hbox{\vrule height 0.85\b@gheight
                                depth 0.35\b@gheight width\z@ }}
\def\rm{\n@expand\f@m0 }
\def\mit{\n@expand\f@m1 }         \let\oldstyle=\mit
\def\cal{\n@expand\f@m2 }
\def\it{\n@expand\f@m\itfam}
\def\sl{\n@expand\f@m\slfam}
\def\bf{\n@expand\f@m\bffam}
\def\tt{\n@expand\f@m\ttfam}
\def\caps{\n@expand\f@m\cpfam}    
\def\em@{\rel@x\ifnum\f@ntkey=0 \it \else
        \ifnum\f@ntkey=\bffam \it \else \rm \fi \fi }
\def\em{\n@expand\em@}
\def\fourteenpoint{\fourteenf@nts \samef@nt \b@gheight=14pt \setstr@t }
\def\twelvepoint{\twelvef@nts \samef@nt \b@gheight=12pt \setstr@t }
\def\tenpoint{\tenf@nts \samef@nt \b@gheight=10pt \setstr@t }
\normalbaselineskip = 19.2pt plus 0.2pt minus 0.1pt 
\normallineskip = 1.5pt plus 0.1pt minus 0.1pt
\normallineskiplimit = 1.5pt
\newskip\normaldisplayskip
\normaldisplayskip = 14.4pt plus 3.6pt minus 10.0pt 
\newskip\normaldispshortskip
\normaldispshortskip = 6pt plus 5pt
\newskip\normalparskip
\normalparskip = 6pt plus 2pt minus 1pt
\newskip\skipregister
\skipregister = 5pt plus 2pt minus 1.5pt
\newif\ifsingl@
\newif\ifdoubl@
\newif\iftwelv@  \twelv@true
\def\singlespace{\singl@true\doubl@false\spaces@t}
\def\doublespace{\singl@false\doubl@true\spaces@t}
\def\normalspace{\singl@false\doubl@false\spaces@t}
\def\Tenpoint{\tenpoint\twelv@false\spaces@t}
\def\Twelvepoint{\twelvepoint\twelv@true\spaces@t}
\def\spaces@t{\rel@x
      \iftwelv@ \ifsingl@\subspaces@t3:4;\else\subspaces@t1:1;\fi
       \else \ifsingl@\subspaces@t3:5;\else\subspaces@t4:5;\fi \fi
      \ifdoubl@ \multiply\baselineskip by 5
         \divide\baselineskip by 4 \fi }
\def\subspaces@t#1:#2;{
      \baselineskip = \normalbaselineskip
      \multiply\baselineskip by #1 \divide\baselineskip by #2
      \lineskip = \normallineskip
      \multiply\lineskip by #1 \divide\lineskip by #2
      \lineskiplimit = \normallineskiplimit
      \multiply\lineskiplimit by #1 \divide\lineskiplimit by #2
      \parskip = \normalparskip
      \multiply\parskip by #1 \divide\parskip by #2
      \abovedisplayskip = \normaldisplayskip
      \multiply\abovedisplayskip by #1 \divide\abovedisplayskip by #2
      \belowdisplayskip = \abovedisplayskip
      \abovedisplayshortskip = \normaldispshortskip
      \multiply\abovedisplayshortskip by #1
        \divide\abovedisplayshortskip by #2
      \belowdisplayshortskip = \abovedisplayshortskip
      \advance\belowdisplayshortskip by \belowdisplayskip
      \divide\belowdisplayshortskip by 2
      \smallskipamount = \skipregister
      \multiply\smallskipamount by #1 \divide\smallskipamount by #2
      \medskipamount = \smallskipamount \multiply\medskipamount by 2
      \bigskipamount = \smallskipamount \multiply\bigskipamount by 4 }
\def\normalbaselines{ \baselineskip=\normalbaselineskip
   \lineskip=\normallineskip \lineskiplimit=\normallineskip
   \iftwelv@\else \multiply\baselineskip by 4 \divide\baselineskip by 5
     \multiply\lineskiplimit by 4 \divide\lineskiplimit by 5
     \multiply\lineskip by 4 \divide\lineskip by 5 \fi }
\Twelvepoint  %
\interlinepenalty=50
\interfootnotelinepenalty=5000
\predisplaypenalty=9000
\postdisplaypenalty=500
\hfuzz=1pt
\vfuzz=0.2pt

\def\noblackbox{\overfullrule=0pt}
\noblackbox

\newcount\infootnote
\infootnote=0
\def\foot#1#2{\infootnote=1
\footnote{${}^{#1}$}{\vtop{\baselineskip=.75\baselineskip
\advance\hsize by -\parindent\noindent{\xrm #2}}}\infootnote=0$\,$}
\newcount\refcount
\refcount=1
\newwrite\refwrite
\def\oldsize{\ifnum\infootnote=1\xold\else\old\fi}
\def\ref#1#2{
	\def#1{{{\oldsize\the\refcount}}\ifnum\the\refcount=1\immediate\openout\refwrite=\jobname.refs\fi\immediate\write\refwrite{\item{[{\xold\the\refcount}]} 
	#2\hfill\par\vskip-2pt}\xdef#1{{\noexpand\oldsize\the\refcount}}\global\advance\refcount by 1}
	}
\def\refout{\catcode`\@=11
        \xrm\immediate\closeout\refwrite
        \vskip.6\baselineskip
        {\noindent\twelvebf References}\hfill\vskip.2\baselineskip
        \baselineskip=.75\baselineskip
        \input\jobname.refs
        \baselineskip=4\baselineskip \divide\baselineskip by 3
        \catcode`\@=\active\rm}

\def\hepth#1{\href{http://xxx.lanl.gov/abs/hep-th/#1}{hep-th/{\xold#1}}}
\def\jhep#1#2#3#4{\href{http://jhep.sissa.it/stdsearch?paper=#2\%28#3\%29#4}{J. High Energy Phys. {\xbold #1#2} ({\xold#3}) {\xold#4}}}
\def\AP#1#2#3{Ann. Phys. {\xbold#1} ({\xold#2}) {\xold#3}}
\def\ATMP#1#2#3{Adv. Theor. Math. Phys. {\xbold#1} ({\xold#2}) {\xold#3}}
\def\CMP#1#2#3{Commun. Math. Phys. {\xbold#1} ({\xold#2}) {\xold#3}}
\def\CQG#1#2#3{Class. Quantum Grav. {\xbold#1} ({\xold#2}) {\xold#3}}

\def\JHEP{\jhep}

\def\MPLA#1#2#3{Mod. Phys. Lett. {\xbf A}{\xbold#1} ({\xold#2}) {\xold#3}}
\def\NPB#1#2#3{Nucl. Phys. {\xbf B}{\xbold#1} ({\xold#2}) {\xold#3}}

\def\PLB#1#2#3{Phys. Lett. {\xbf B}{\xbold#1} ({\xold#2}) {\xold#3}}

\def\PRD#1#2#3{Phys. Rev. {\xbf D}{\xbold#1} ({\xold#2}) {\xold#3}}

\newcount\sectioncount
\sectioncount=0
\def\section#1#2{\global\eqcount=0
	\global\subsectioncount=0
        \global\advance\sectioncount by 1
	\ifnum\sectioncount>1
	        \vskip\baselineskip
	\fi
	\noindent
        \line{\twelvebf\the\sectioncount. #2\hfill}
		\vskip.2\baselineskip\noindent
        \xdef#1{{\old\the\sectioncount}}}
\newcount\subsectioncount
\def\subsection#1#2{\global\advance\subsectioncount by 1
	\vskip.6\baselineskip\noindent
	\line{\twelvecp\the\sectioncount.\the\subsectioncount. #2\hfill}
	\vskip.5\baselineskip\noindent
	\xdef#1{{\old\the\sectioncount}.{\old\the\subsectioncount}}}
\newcount\appendixcount
\appendixcount=0
\def\appendix#1{\global\eqcount=0
        \global\advance\appendixcount by 1
        \vskip2\baselineskip\noindent
        \ifnum\the\appendixcount=1
        \hbox{\twelvecp Appendix A: #1\hfill}\vskip\baselineskip\noindent\fi
    \ifnum\the\appendixcount=2
        \hbox{\twelvecp Appendix B: #1\hfill}\vskip\baselineskip\noindent\fi
    \ifnum\the\appendixcount=3
        \hbox{\twelvecp Appendix C: #1\hfill}\vskip\baselineskip\noindent\fi}

\newcount\eqcount
\eqcount=0
\def\Eqn#1{\global\advance\eqcount by 1
\ifnum\the\sectioncount=0
	\xdef#1{{\oldsize\the\eqcount}}
	\eqno({\oldstyle\the\eqcount})
\else
        \ifnum\the\appendixcount=0
	        \xdef#1{{\old\the\sectioncount}.{\old\the\eqcount}}
                \eqno({\oldstyle\the\sectioncount}.{\oldstyle\the\eqcount})\fi
        \ifnum\the\appendixcount=1
	        \xdef#1{{\oldstyle A}.{\old\the\eqcount}}
                \eqno({\oldstyle A}.{\oldstyle\the\eqcount})\fi
        \ifnum\the\appendixcount=2
	        \xdef#1{{\oldstyle B}.{\old\the\eqcount}}
                \eqno({\oldstyle B}.{\oldstyle\the\eqcount})\fi
        \ifnum\the\appendixcount=3
	        \xdef#1{{\oldstyle C}.{\old\the\eqcount}}
                \eqno({\oldstyle C}.{\oldstyle\the\eqcount})\fi
\fi}
\def\Fqn#1#2{\global\advance\eqcount by 1
\ifnum\the\sectioncount=0
	\xdef#1{{\old\the\eqcount}}
	\xdef#2{{\xold\the\eqcount}}
	\eqno({\oldstyle\the\eqcount})
\else
	\xdef#1{{\old\the\sectioncount}.{\old\the\eqcount}}
	\xdef#2{{\xold\the\sectioncount}.{\xold\the\eqcount}}
	\eqno({\oldstyle\the\sectioncount}.{\oldstyle\the\eqcount})
\fi}
\def\eqn{\global\advance\eqcount by 1
\ifnum\the\sectioncount=0
	\eqno({\oldstyle\the\eqcount})
\else
        \ifnum\the\appendixcount=0
                \eqno({\oldstyle\the\sectioncount}.{\oldstyle\the\eqcount})\fi
        \ifnum\the\appendixcount=1
                \eqno({\oldstyle A}.{\oldstyle\the\eqcount})\fi
        \ifnum\the\appendixcount=2
                \eqno({\oldstyle B}.{\oldstyle\the\eqcount})\fi
        \ifnum\the\appendixcount=3
                \eqno({\oldstyle C}.{\oldstyle\the\eqcount})\fi
\fi}
\def\multi{\global\advance\eqcount by 1}
\def\multieq#1#2{\xdef#1{{\old\the\eqcount#2}}
        \eqno{({\oldstyle\the\eqcount#2})}}
\newtoks\url
\def\Href#1#2{\catcode`\#=12\url={#1}\catcode`\#=\active#2}
\def\href#1#2{{#2}}

\hsize=16cm
\vsize=24.1cm
\frenchspacing
\footline={}
\raggedbottom

\def\ss{\scriptstyle}

\def\*{\partial}
\def\punkt{\,\,.}
\def\komma{\,\,,}

\def\={\!=\!}
\def\small#1{{\hbox{$#1$}}}

\def\fraction#1{\small{1\over#1}}
\def\fr{\fraction}
\def\Fraction#1#2{\small{#1\over#2}}
\def\Fr{\Fraction}
\def\tr{\hbox{\rm tr}}
\def\eg{{\it e.g.}}

\def\ie{{\it i.e.}}

\def\nlni{\hfill\break}

\def\d{\delta}
\def\e{\varepsilon}
\def\g{\gamma}
\def\l{\lambda}

\def\R{{\Bbb R}}
\def\H{{\Bbb H}}

\def\Re{\hbox{Re}\,}

%
%

\def\m{\mu}
\def\n{\nu}

\def\r{\varrho}

\def\e{\varepsilon}

\def\d{\partial}

\def\R{{\Bbb R}}

\def\H{{\Bbb H}}

\def\rightarrowover#1{\raise9pt\vtop{\baselineskip=0pt\lineskip=0pt
      \ialign{##\cr$\rightarrow$ \cr
                \hfill $#1$\hfill\cr}}}
\def\leftarrowover#1{\raise9pt\vtop{\baselineskip=0pt\lineskip=0pt
      \ialign{##\cr$\leftarrow$ \cr
                \hfill $#1$\hfill\cr}}}
		
\def\Re{\hbox{Re}}
\def\Im{\hbox{Im}}

%
%
%

\centerline{\sixteenhelvbold M5-Branes and Matrix Theory} 

\vskip1cm

\centerline{\twelvehelvbold Martin Cederwall}
\centerline{\twelvehelvbold Henric Larsson}
\vskip.7cm

{\Tenpoint
\centerline{\tenit Department of Theoretical Physics}
\centerline{\tenit G\"oteborg University 
and Chalmers University of Technology }
\centerline{\tenit S-412 96 G\"oteborg, Sweden}
}

\vskip1cm

{\narrower\noindent\Tenpoint
\underbar{Abstract:} We consider super-membranes ending on
M5-branes, with the aim of deriving the appropriate matrix theories
describing different situations. Special attention is given to the
case of non-vanishing (selfdual) $C$-field. We identify the relevant
deformation of the six-dimensional super-Yang--Mills theory whose
dimensional reduction is the matrix theory for membranes in the
presence of M5-branes. Possible applications and limitations of the
models are discussed.
\smallskip}

\vskip1cm


\ref\CederwallLarsson{M. Cederwall and H. Larsson, {\xit to appear}.}

\ref\OMTheory{R. Gopakumar, S. Minwalla, N. Seiberg and A. Strominger,
{\xit ``OM theory in diverse dimensions''}, \jhep{00}{08}{2000}{008}
[\hepth{0006062}];
\nlni E. Bergshoeff, D.S. Berman, J.P. van der Schaar and  P. Sundell,
{\xit ``Critical fields on the M5-brane and noncommutative open
strings''}, \PLB{492}{2000}{193} [\hepth{0006112}].}

\ref\IndexThm{S. Sethi and M. Stern, {\xit ``The structure of the D0-D4
bound state''}, \NPB{578}{2000}{163} [\hepth{0002131}].}

\ref\CederwallStretched{M. Cederwall, {\xit ``Open and winding membranes,
affine matrix theory and matrix string theory''}, 
\JHEP{02}{12}{2002}{005} [\hepth{0210152}].}

\ref\AharonyEtAl{O. Aharony, M. Berkooz, S. Kachru, N. Seiberg 
and E. Silverstein, {\xit ``Matrix description of interacting 
theories on six dimensions''}, \ATMP{1}{1998}{148} [\hepth{9707079}];\nlni
O. Aharony, M. Berkooz and N. Seiberg, {\xit ``Light-cone description of (2,0)
superconformal theories in six dimensions''},
\ATMP{2}{1998}{119} [\hepth{9712117}].}

\ref\Berkooz{Ori J. Ganor and Joanna L. Karczmarek,
{\xit ``M(atrix)-theory scattering in the noncommutative (2,0) theory''},
\JHEP{00}{10}{2000}{024} [\hepth{0007166}];
\nlni M. Berkooz, {\xit ``Light-like (2,0) noncommutativity and
light-cone rigid open membrane theory''}, \hepth{0010158}.}

\ref\deWitHoppeNicolai{B. de Wit, J. Hoppe and H. Nicolai,
	{``On the quantum mechanics of supermembranes''},
	\NPB{305}{1988}{545}.}

\ref\deWitLuscherNicolai{B. de Wit, M L\"uscher and H. Nicolai,
	{\xit ``The supermembrane is unstable''},
	\NPB{320}{1989}{135}.}

\ref\deWitMarquardNicolai{B. de Wit, U. Marquard and H. Nicolai,
	{\xit ``Area preserving diffeomorphisms and supermembrane 
	Lorentz invariance''}, \CMP{128}{1990}{39}.}

\ref\CederwallOpenMembrane{M. Cederwall, 
	{\xit ``Boundaries of {\xold11}-dimensional membranes''}, 
	\MPLA{12}{1997}{2641} [\hepth{9704161}].}

\ref\HoravaWitten{P. Ho\v rava and E. Witten,
	{\xit ``Heterotic and type I string dynamics from eleven-dimensions''},
	\NPB{460}{1996}{506} [\hepth{9510209}];
	{\xit ``Eleven-dimensional supergravity on a manifold with boundary''},
	\NPB{475}{1996}{94} [\hepth{9603142}].}

\ref\FairlieZachos{D.B. Fairlie and C.K. Zachos, 
	{\xit ``Infinite dimensional algebras, sine brackets and 
	{\xit su}($\ss\infty$)''}, \PLB{224}{1989}{101}.}

\ref\KimRey{N. Kim and S.-J. Rey,
	{\xit ``M(atrix) theory on an orbifold and twisted membrane''},
	\NPB{504}{1997}{189} [\hepth{9701139}].}

\ref\EzawaMatsuoMurakami{K. Ezawa, Y. Matsuo and K. Murakami,
	{\xit ``Matrix regularization of open supermembrane: 
	towards M theory five-brane via open supermembrane''},
	\PRD{57}{1998}{5118} [\hepth{9707200}].}

\ref\BeckerBecker{K. Becker and M. Becker, {\xit ``Boundaries in M theory''},
	\NPB{472}{1996}{221} [\hepth{9602071}].}

\ref\Hoppe{J. Hoppe, {\xit ``Zero energy states in supersymmetric 
matrix models''},
	\CQG{17}{2000}{1101}, and references therein.}

\ref\MatrixReviews{M.J. Duff, {\xit ``Supermembranes''}, \hepth{9611203};
	\nlni H. Nicolai and R. Helling, 
		{\xit ``Supermembranes and (M)atrix theory''}, \hepth{9809103};
	\nlni B. de Wit, {\xit ``Supermembranes and super matrix models''}, 
		\hepth{9902051}; 
	\nlni T. Banks, {\xit ``TASI lectures on matrix theory''}, 
	\hepth{9911068};
	\nlni W. Taylor, {\xit ``(M)atrix theory: Matrix quantum mechanics 
		as fundamental theory''}, \hepth{0101126}.}

\ref\deWitPeetersPlefka{B. de Wit, K. Peeters and J. Plefka,
	{\xit ``Supermembranes with winding''}, 
	\PLB{409}{1997}{117} [\hepth{9705225}]}

\ref\BFSS{T. Banks, W. Fischler, S.H. Shenker and L. Susskind,
	{\xit ``M theory as a matrix model: a conjecture''}, 
	\PRD{55}{1997}{5112} [\hepth{9610043}].}

\ref\SethiStern{S. Sethi and M. Stern, {\xit ``D-brane bound states redux''},
		\CMP{194}{1998}{675} [\hepth{9705046}].}

\ref\KacSmilga{V.G. Kac and A.V. Smilga, 
		{\xit ``Normalized vacuum states in N=4 supersymmetric 
		Yang--Mills quantum mechanics with any gauge group''},
		\NPB{571}{2000}{515} [\hepth{9908096}].}

\ref\BST{E. Bergshoeff, E. Sezgin and P.K. Townsend,
	{\xit ``Supermembranes and eleven-dimensional supergravity''},
	\PLB{189}{1987}{75}; 
	{\xit ``Properties of the eleven-dimensional super membrane theory''},
	\AP{185}{1988}{330}.}

\ref\BSTT{E. Bergshoeff, E. Sezgin, Y. Tanii and P.K. Townsend,
	{\xit ``Super {\tenit p}-branes as gauge theories of 
	volume preserving diffeomorphisms''},
	\AP{199}{1990}{340}.}

\ref\ChuSezgin{C.-S. Chu and E. Sezgin, 
	{\xit ``M five-brane from the open supermembrane''},
	\JHEP{97}{12}{1997}{001} [\hepth{9710223}].}

\ref\FuchsSchweigert{J. Fuchs and C. Schweigert, 
	{\xit ``Symmetries, Lie algebras and representations''},
	Cambridge University Press, 1997.}

\ref\DVV{R. Dijkgraaf, E. Verlinde and H. Verlinde,
	{\xit ``Matrix string theory''}, \NPB{500}{1997}{43} 
	[\hepth{9703030}].} 

\ref\SekinoYoneya{Y. Sekino and T. Yoneya, 
	{\xit ``From supermembrane to matrix string''},
	\NPB{619}{2001}{22} [\hepth{0108176}].}

\ref\Minic{Dj. Mini\'c, {\xit ``M-theory and deformation quantization''},
	\hepth{9909022}.}	

\ref\BraxMourad{Ph. Brax and J. Mourad, 
	{\xit ``Open supermembranes in eleven-dimensions''}
	\PLB{408}{1997}{142} [\hepth{9704165}];
	{\xit ``Open supermembranes coupled to M theory five-branes''},
	\PLB{416}{1998}{295} [\hepth{9707246}].}

\ref\Henningson{A. Gustavsson and M. Henningson, 
	{\xit ``A short representation of the six-dimensional (2,0) algebra''},
	\JHEP{01}{06}{2001}{054} [\hepth{0104172}].}

\ref\BermanEtAl{D.S. Berman, M. Cederwall, U. Gran, H. Larsson,
      M. Nielsen, B.E.W. Nilsson and P. Sundell,
{\xit ``Deformation independent open brane metrics and generalized
      theta parameters''},
	\jhep{02}{02}{2002}{012} [\hepth{0109107}].}

\ref\Schaar{J.P. van der Schaar, {\xit ``The reduced open membrane metric''},
	\jhep{01}{08}{2001}{048} [\hepth{0106046}].}

\ref\BergshoeffNonComm{E. Bergshoeff, D.S. Berman, J.P. van der Schaar
and P. Sundell, {\xit ``A noncommutative M-theory five-brane''},
\NPB{590}{2000}{173} [\hepth{0005026}].}

\section\intro{Introduction}Supermembrane [\BST] theory [\deWitHoppeNicolai]
is a very promising
candidate for a microscopic description of M-theory. Although it is not
background invariant, it gives a completely new picture of the nature of
space and time at small scales, together with a description of 
quantum-mechanical states that goes beyond local quantum field theory.
These features are most clear in the matrix [\BFSS] truncation 
[\deWitHoppeNicolai] of the membrane.
It is widely appreciated that first-quantised supermembrane theory 
through its continuous spectrum [\deWitLuscherNicolai] is capable of
describing an entire (``multi-particle'') Fock space. 
For reviews on the subject of membranes and matrices, see ref. 
[\MatrixReviews].
Due to the immense
technical difficulties associated with actual calculations in the theory, 
which is non-linear and inherently non-perturbative, few quantitative
features are known in addition to the general picture, which is supported 
by many qualitative arguments. The maybe most important one is the proof
that $su(N)$ matrix theory has a unique supersymmetric ground state
[\SethiStern,\KacSmilga], which gives
the relation to the massless degrees of freedom of $D=11$ supergravity.

Many situations in M-theory backgrounds involve membranes that are not closed.
Supermembranes may end on ``defects'', \ie, {\old5}-branes and {\old9}-branes
[\HoravaWitten,\CederwallOpenMembrane,\BraxMourad,\ChuSezgin,\BeckerBecker,\KimRey,\CederwallStretched].
It is urgent to have some mathematical formulation of these situations
in order to understand the microscopic properties of physics in such
backgrounds. One old enigma is the nature of the theory on multiple
{\old5}-branes, which we address in the present talk.
There are several issues to be resolved. The membrane may be stretched
between multiple {\old5}-branes, and the truncation has to be
consistent with this situation. In addition, the $C$-field, the
{\old3}-form potential of M-theory, may take some non-vanishing
selfdual value on the {\old5}-brane. The new results contained in this
talk refer to the latter situation. There are several reasons to
consider this specific situation. It should be connected to the theory
on multiple M{\old5}-branes, which is some kind of non-abelian theory
of selfdual tensors [\Henningson]. It should be possible to verify the
decoupling limit of OM-theory [\OMTheory] from microscopic
considerations. There might also be information about the open
membrane metric [\Schaar,\BermanEtAl] and maybe even some clue 
concerning the proper
generalisation of the string endpoint non-commutativity to membranes 
[\BergshoeffNonComm,\BermanEtAl].

We start out by reviewing the consistent truncation of membranes to
matrices via non-commutativity in section {\old2}. Section {\old3}
describes how this construction is generalised to situations where the
membrane has a boundary 
[\deWitPeetersPlefka,\EzawaMatsuoMurakami,\CederwallStretched,\SekinoYoneya]. 
Here we review the alternative constructions
present in the literature, and discuss their relative
applicability. In section {\old4}, we generalise the picture to
include non-vanishing $C$-field, both light-like 
[\Berkooz] and general. We
identify the deformation of the {\old6}-dimensional super-Yang--Mills
theory whose dimensional reduction is the matrix theory associated to
turning on the $C$-field. In section {\old5}, we discuss the possible
applications and limitations of the model.

\section\membranestomatrices{From membranes to matrices}We start
from the action for the supermembrane coupled to an on-shell
background of $D=11$ supergravity,
$$
S=-T\int d^3\xi\sqrt{-g}+T\int C\punkt\Eqn\MembraneAction
$$
Here, the metric and $C$-field are pullbacks from superspace to the
bosonic world-volume.
In what follows, we will consider flat backgrounds, but allow for
non-zero constant $C$.

Let us first remind of the consistent truncation to matrix theory of a
closed membrane (we just display the bosonic degrees of freedom;
fermions are straightforwardly included). Here, the $C$-field is irrelevant.
In light-cone gauge, where reparametrisation invariance is used up
except for area-preserving diffeomorphisms of the membrane ``space-sheet''.
The light-cone hamiltonian $p^-$ is given by
$$
{p^+p^-\over A}=\int
d^2\xi\left(P_IP_I+{T^2\over4}\{X^I,X^J\}\{X^I,X^J\}\right)
\komma\eqn
$$
where $A$ is the parametric area of the space-sheet, and
$\{A,B\}=\e^{ij}\*_iA\*_jB$ is the ``Poisson bracket'' on the
space-sheet. The remaining gauge invariance is generated by the
Poisson bracket as $\delta_fA=\{f,A\}$ [\BSTT]. Even though it is known that
the algebra of area-preserving diffeomorphisms in a certain sense is 
$su(\infty)$ [\deWitHoppeNicolai,\FairlieZachos], $su(N)$ is not
contained as a subalgebra, and there is no way of getting to $su(N)$
matrix theory as a consistent truncation. 

In order to obtain matrix theory as a consistent truncation, one
introduces a non-commutativity on the membrane space-sheet (for
simplicity, we consider a toroidal membrane), 
$[\xi^1,\xi^2]=\theta$, encoded in the Weyl-ordered star product
$
f\star
g=f\exp({i\over2}\theta\e^{ij}\leftarrowover{\*_i}\rightarrowover{\*_j})g
$.
Commutators between Fourier modes 
become
$
[e^{ik\cdot\xi},e^{ik'\cdot\xi}] =-2i\sin({\theta\over2}\e^{ij}k_ik'_j)\,
e^{i(k+k')\cdot\xi}
$.
The Poisson bracket is recovered as $\{\cdot,\cdot\}
=-i\lim_{\theta\rightarrow0}\theta^{-1}[\cdot,\cdot]$. Choosing
$\theta={2\pi\over N}$ implies that the functions $e^{iN\xi^1}$,
$e^{iN\xi^2}$ are central. Their action on any function can
consistently be modded out according to the equivalence relation
$e^{iN\xi^{1,2}}\star f\approx f$. The remaining ``square of
functions'' with mode numbers ranging from 0 to $N-1$
generate $u(N)$ [\deWitHoppeNicolai,\FairlieZachos].

The model thus obtained as a consistent truncation of the
supermembrane is an $su(N)$ supersymmetric matrix model, identical to
the dimensional reduction to $D=1$ of $D=10$ super-Yang--Mills
theory. This example sets the procedure we want to apply to other
cases: deform by non-commutativity, replace Poisson brackets by
commutators and perform a consistent truncation of the deformed
theory.
We would like to stress the importance of making a consistent
truncation, in contrast to an approximation; the fact that the
commutator obeys the same algebraic identities as the Poisson bracket
means that one has control over the symmetries of the model, \eg\
supersymmetry. The only symmetries that are lost in the matrix
truncation are the super-Poincar\'e generators that are non-linearly
realised in the light-cone gauge.


\section\membraneswithboundary{Matrix theory for membranes with
boundary}Let us now turn to the first modification of the previous
situation, namely when the membrane has boundaries (we think of these
as lying on M{\old5}-branes, but much if what is said applies to any
possible boundary). It is expected that the ``no-topology'' theorem
that applies for closed membranes persists for membranes with
boundary, so that it is irrelevant \eg\ whether a membrane ending on
only one {\old5}-brane is modeled as a half sphere, a half torus or
some more complicated manifold. This is an assumption we make; a proof
would be desirable.

We can distinguish between two classes of approaches to this kind of
configuration:
\item{\bf A.}{This approach was first physically motivated by double
dimensional reduction to a D{\old4}-brane. The theory obtained after
reduction is $D=5$ super-Yang--Mills, and opening up the sixth
direction should correspond to a strong coupling limit. In this limit,
path integrals are dominated by saddle points at the moduli space 
of ``instanton'' solitons. The moduli space of $N$ instantons in
U($k$) SYM has dimension $4kN$. The matrix theory should have this
space as Higgs branch. It is the dimensional reduction of a $D=6$
U($N$) SYM with one adjoint and $k$ fundamental hypermultiplets 
[\AharonyEtAl]. 
We will motivate this from the point of view of the supermembrane. 
}
\item{\bf B.}{For a fixed membrane topology (a half torus, say), the
boundary conditions may be solved, at least when $C=0$ 
(see [\CederwallStretched]). For the 5 directions transverse to the (flat) 
M{\old5}-brane one gets Dirichlet boundary conditions, which for torus
topology means sine functions, and for the 4 transverse (2 have been
eliminated when going to light-c\^one gauge) one gets Neumann boundary
conditions, leading to cosine functions. The sine functions generate
SO($N$) [\KimRey], and the cosine function transform as the symmetric
representation. The matrix model obtained is the dimensional reduction
of a $D=6$ SO($N$) SYM with a hypermultiplet in the symmetric representation.
}

\noindent A couple of comments can be made. The first one concerns the
global symmetries of the matrix theory and of the M-theory
configuration it describes. A $D=6$ SYM theory with hypermultiplets
has a lagrangian
$$
\eqalign{
{\cal L}=&-\fr4F^A_{\m\n}F^{A\m\n}+\fr2\Re(\l^{\dagger A}\g^\m D_\m\l^A)
	-\fr2D^\m\phi^ID_\m\phi^*_I
		-\fr2\Re(\psi^\dagger_I\tilde\g^\m D_\m\psi^I)\cr
	&-(\r^A)_I{}^J\Re(\l^{\dagger A}\psi^I\phi^*_J)
	+\fr8(\r^A)_I{}^J(\rho^A)_K{}^L\phi^I\phi^*_J\phi^K\phi^*_L\punkt\cr
}\Eqn\SYMAction
$$
We use the isomorphism $\hbox{Spin}(1,5)\approx\hbox{SL}(2;\H)$ 
and two-component 
quaternionic spinors. The scalars $\phi$ are quaternionic, $\phi=\phi_ie_i$,
where $i=1,\ldots,4$. 
Indices $I,J,\ldots$ label the representation of 
the hypermultiplet, and $\r$ is the representation matrix.
For real hypermultiplets (as in case {\bf B} above) there is an 
$\hbox{SU}(2)_L\times\hbox{SU}(2)_R$ R-symmetry realised as multiplication
by unit quaternions as
$$
\matrix{A\rightarrow A\komma\hfill\quad&\l\rightarrow\l h_L\komma\hfill\cr
\phi\rightarrow h_L^*\phi h_R\komma\hfill\quad
&\psi\rightarrow\psi h_R\punkt\hfill\cr}
\eqn
$$
When the hypermultiplet is complex, the right action is occupied by
the gauge group. If there is an even number of hypermultiplets in the
same representation, there will however be a flavour symmetry SU($2k$)
The representations
(specified by dimensions) of the fields under the Lorentz rotations
and R-symmetry are
thus 
$$
\matrix{A:(6,1,1)\komma\hfill\quad&\l:(4,2,1)\komma\hfill\cr
\phi:(1,2,2\,\hbox{or}\,1)\komma\hfill\quad
&\psi:(\overline4,1,2\,\hbox{or}\,1)\hfill\cr}
\eqn
$$
(the last alternative for $su(2)_R$ representations of the hypermultiplet 
is for the minimal content of a complex representation).
The R-symmetry of the super-Yang--Mills theory is the rotation
symmetry of the membrane/matrix theory in light-c\^one gauge, and the
SO(5) rotation symmetry remaining on the super-Yang-Mills side after
dimensional reduction is the R-symmetry of the membrane/matrices.

The supersymmetry transformation rules are
$$
\matrix{\delta_\e A^A_\m=\Re(\e^\dagger\g_\m\l^A)\komma\hfill\quad
&\delta_\e\l^A=\fr2F^A_{\m\n}\g^{\m\n}\e+\e W^A\komma\hfill\cr
\delta_\e\phi^I=-\e^\dagger\psi^I\komma\hfill\quad
&\delta_\e\psi^I=\g^\m\e D_\m\phi^I\komma\hfill\cr}
\Eqn\SuperSymmetry
$$
where $\e$ is a spinor in the same representation as $\l$. Since we
later want to identify the presence of a $C$-field with certain
deformations of SYM that leave supersymmetry unbroken, we have written 
the transformation of the adjoint spinor using
$W^A$, an imaginary
quaternion (\ie, transforming in $(1,3,1)$)
in the adjoint of the gauge group. In the undeformed case, 
$W^A=W^A_0=\fr2(\r^A)_I{}^J\phi^I\phi^\star_J$. Note that the hypermultiplet
potential is the square of $W$, $V(\phi)=\fr2W^AW^{\star A}$. The
deformations will be encoded in the form of $W$. The most convenient way of
checking supersymmetry is to note that $W$ is contained in the same
supermultiplet as the hypermultiplet gauge current:
$$
\eqalign{
\delta_\e W^A&=-\Im(\e^\dagger\mu^A)\komma\cr
\delta_\e\mu^A&=J_\mu^A\g^\mu\e+\g^\mu\e D_\mu W^A\komma\cr
}\eqn
$$
where $\mu^A=\mu_0^A=(\r^A)_I{}^J\psi^I\phi^\star_J$, 
$J_\mu^A={1\over2}(\r^A)_I{}^J(D_\mu\phi^I\phi^\star_J
        -\phi^ID_\mu\phi^\star_J-\psi^\dagger_J\tilde\g_\mu\psi^I)$.

Before turning to the derivation of case {\bf A} from the supermembrane,
let us discuss the advantages and limitations of the two approaches
and some aspects of their physical content.
Both cases are defined as dimensional reductions of $D=6$ SYM with
matter. The expression for the
potential is a sum of positive semidefinite terms, so the Higgs branch
is determined by $W=0$. In light of the correspondence with
five-dimensional physics mentioned above, it is interesting to
investigate the geometry of the Higgs branch. The low-energy limit of
adiabatic motion on the Higgs branch is also the situation when bulk
excitations (gravity) decouple. Counting the dimension of the Higgs
branch as $\#(\hbox{scalar matter fields})-\#W-\hbox{dim(gauge
group)}$,
one gets in case {\bf A}: $4N^2+4kN-3N^2-N^2=4kN$, 
and in case {\bf B}:
$4{N(N+1)\over2}-3{N(N-1)\over2}-{N(N-1)\over2}=4N$.
Closer investigation reveals that the spaces agree for $k=1$, and
that the Higgs branch then is
$\R^4\times({R^{4(N-1)}/P_N})$, interpreted as the space of loci of N
indistinguishable partons/D0-branes.
This is a flat hyper-K\"ahler space with
conical singularities where partons coincide, which is where the Higgs
branch intersects the Coulomb branch.

There is an index theorem [\IndexThm] 
stating that the matrix theory has a unique supersymmetric ground
state. The eight fermion zero modes lie in the representation $(4,1,2)$, so 
the ground state is the breaking to SO(5)$\times$SU(2) of an SO(8)
spinor $8_s\oplus8_c$ when the vector decomposes as
$8_v\rightarrow(4,2)$ (the ``Hopf breaking'').
Then $8_s\rightarrow(1,3)\oplus(5,1)$ and $8_c\rightarrow(4,2)$,
giving the bosonic and fermionic fields of the selfdual $D=6$ tensor
multiplet in the light-c\^one gauge.

Note that approach {\bf B} does not seem to accommodate multiple
M{\old5}-branes in a natural way. On the other hand, approach {\bf A},
as we will see, is less adaptable to incorporate the stringy nature of
the membrane boundary. This is connected to the way it is derived from
the membrane below; no boundary conditions are solved, the nature of
the boundary is rather point-like. It is also unclear how {\bf A}
generalises to separated M{\old5}-branes. Concerning the
incorporation of a non-vanishing $C$-field (following section),
approach {\bf A} has the advantage of being more or less directly
applicable, while approach {\bf B} encounters problems, due to the
difficulty (impossibility?) of solving the boundary conditions in the
presence of a $C$-field.

Let us sketch briefly how case {\bf A} is derived as a consistent
truncation from the supermembrane. As we already mentioned, no
boundary conditions are solved before performing the matrix
truncation. Instead we introduce the ``boundary'' through the
truncated $\delta$-function 
$
\Delta\equiv\sum_{n=0}^{N-1}e^{in\r}
$ 
(we consider a boundary located at $\r=0$, where $\r$ for simplicity is a
coordinate on a torus).
Due to the identities $\Delta^2=N\Delta$ and 
$\Delta\star f\star\Delta=0$, left and right star multiplication with 
$\Delta/\sqrt N$ projects on two ``boundary representations'' $N$ and
$\bar N$ with opposite U(1) charge under adjoint action of $\Delta$,
\eg, $[\Delta,\Delta\star f]=\Delta\star\Delta\star f-\Delta\star
f\star\Delta=N\Delta\star f$. Introduction of the ``boundary'' breaks
$su(N)$ to $su(N-1)\oplus u(1)$. Higher rank $\delta$-functions (sums of
$\Delta$'s with $\r$ shifted by $2\pi\over N$ times an integer) gives 
$su(N-k)\oplus su(k)\oplus u(1)$. 

Let us also show how approach {\bf A} generalises to a situation where
the membrane is stretched between two separated parallel
M{\old5}-branes (separation $L/2$) 
or where the membrane is wound on a non-contractible
circle (length $L$) [\CederwallStretched].
The mode expansion of a coordinate of a cylindrical membrane in the
separation direction then contains a linear
term in addition to the oscillators:
$
Y(\sigma,\r)={L\over2\pi}\r+{1\over\pi}\sum_{n=-\infty}^\infty
\sum_{m=1}^\infty y_{nm}e^{in\sigma}\sin m\r
$.
The star-adjoint action of $\r\over 2\pi$ is identical to 
$-{i\over N}{\d\over\d\sigma}$. $\r\over 2\pi$ is an outer
derivation on the algebra of functions, and its presence means that it
is not consistent to truncate in the $\sigma$-direction. Truncating in
the $\r$-direction only leads to an affine SO($N$) algebra. The matrix
theory is an ``affine matrix theory'', or a matrix string theory,
which is the dimensional reduction to $D=2$ of a $D=6$ SYM theory with
a hypermultiplet in the symmetric representation. Note that the
coupling constant is relevant, since there is a dimensionless quotient
between the eleven-dimensional Planck length and the brane separation.

\section\cfield{Non-vanishing $C$-field}We now turn to the situation
where there is a non-vanishing $C$-field on the M{\old5}-brane (the
gauge invariant statement is in terms of the selfdual {\old3}-form
field strength $F_{(3)}=dB_{(2)}-C_{(3)}$ on the brane). In the
process of choosing a light-c\^one gauge for the membrane, the same is
done for the $C$-field. We choose $C_{-ij}=0$. Then a selfdual $C$
falls in either of the two classes, modulo choice of frame:
\item{1.}{``Light-like'': $C_{ijk}=0$, $C_{+-i}=0$, 
$C_{+ij}$ selfdual in four dimensions. The transverse rotation are
broken as $so(4)\approx su(2)\oplus su(2)\rightarrow su(2)\oplus u(1)$.}
\item{2.}{``Space-like'': $C_{+ij}=0$,
$C_{ijk}=\e_{ijkl}C_{+-l}$. Transverse rotations broken as 
$so(4)\approx su(2)\oplus su(2)\rightarrow (su(2))_{\hbox{\eightrm
diag}}$.}

\noindent We now have to include the Wess--Zumino term of 
eq. (\MembraneAction) in the canonical analysis. The light-c\^one
membrane hamiltonian becomes
$$
\eqalign{
H&\equiv{p^+p^-\over A}=\int d^2\xi\Bigl[\,
\fr2\Pi_I\Pi_I+\Fr{T^2}4\{X^I,X^J\}\{X^I,X^J\}\cr
&-\Fr{T^2}2C_{+-J}C_{+-K}\{X^I,X^J\}\{X^I,X^K\}
-\Fr{T^2}2C_{IJK}C_{+-L}\{X^I,X^J\}\{X^K,X^L\}\cr
&-\Fr{p^+T}{2A}C_{+IJ}\{X^I,X^J\}
\,\Bigr]\komma\cr
}
\Eqn\HamiltonianWithC
$$
where $\Pi_I\,(=\dot X_I)=P_I-\Fr T2C_{IJK}\{X^J,X^K\}-TC_{+-J}\{X_I,X^J\}$.
In order to identify the connection with SYM, it is useful to form the
lagrangian 
$$
L
=\int d^2\xi\Bigl[\,\fr2\dot X^I\dot X^I
+\Fr T2C_{IJK}\dot X^I\{X^J,X^K\}
+TC_{+-J}\dot X^I\{X^I,X^J\}
-V(X)\,\Bigr]
\Eqn\LagrangianWithC
$$
Due to the difficulties with solving the non-linear boundary
conditions in the presence of a $C$-field, we choose to work in the
approach {\bf A}. The light-like case is much simpler, and already
well known (although not, to our knowledge, derived from the membrane).
There, the last line in eq. (\HamiltonianWithC) represents the only
deformation. We note that in the membrane hamiltonian a term
$
\int d\sigma d\r\{A,B\}=\left(\int_{\r=0}-\int_{\r=\pi}\right)
d\sigma A\d_\sigma B
$
is a cocycle that is not well defined in the matrix truncation (since
it is defined using the derivation $\r$.
Any boundary term should be represented by a cocycle, defined by a
derivation $\d$ as $\tr(A[\d,B])$. 
Since a finite-dimensional Lie algebra only has inner
derivations, one may be lead to conclude that it is necessary to use
the affine matrix theory mentioned earlier. This is however not
true. The relevant derivation is the truncated $\delta$-function
$[\Delta,\cdot]$, 
which is inner, so that the
cocycle $\tr(A,[\Delta,B])$ is exact in the space of functions. 
We get two equivalent pictures,
one with a deformed algebra $[A,B]_k=[A,B]+k\tr(A,[\Delta,B])$, and
one with an undeformed algebra (obtained from the deformed generators
by a redefinition containing $\Delta$) and a modified trace involving
$tr\Delta\neq0$. This gives a coupling containing the boundary
representations $N$ and $\bar N$. It amounts to the introduction in
the SYM theory of a Fayet--Iliopoulos term [\Berkooz] by 
$W^A=W_0^A+\zeta^A$, $\zeta$ being a fixed vector in the U(1)
direction defined by $\Delta$. It breaks the rotational $so(4)$
symmetry to $su(2)\oplus u(1)$ and leaves supersymmetry unbroken. Its
effect is to resolve the singularities of the Higgs branch.

Turning to space-like $C$-field, we use the selfduality 
condition\foot{\dagger}{We use a linear 
self-duality condition, although the
self-duality on an M{\xold5}-brane should really be non-linear. It is
not obvious to us why a linear relation seems to produce the right
result.} 
on $C$ to rewrite the terms in the lagrangian (\LagrangianWithC) linear
in time derivatives as
$
\fr2C_{+-J}\bigl(2\dot X^I\{X^I,X^J\}+\e_{KLMJ}\dot
X^K\{X^L,X^M\}\bigr)
$.
Choosing a basis where $C_{+-4}=\gamma=C_{123}$ and splitting
quaternions in real and imaginary parts with $X^4=\Re X$, this can be
rewritten as proportional to 
$f^{\scr ABC}\Re(\dot X^{\scr A}X^{\scr B}X^{\star{\scr C}})$, where
indices ${\scr A},{\scr B},\ldots$ enumerate the truncated basis of
functions. The only contribution from this term which is not a total
derivative comes from the cocycle mentioned earlier, and the relevant
part is then proportional to $\tr(\dot\phi W_0)$, which
leads to the conclusion that space-like $C$-field corresponds to a
deformation of the SYM theory given by 
$$
W=W_0+\g\Im\dot\phi\komma
\qquad\mu=\mu_0+\g\dot\psi\komma
\eqn
$$ 
where the deformations take values in $u(1)$. Of course, also the
potential terms have to be matched against the SYM theory. It is
straightforward to show, using the supersymmetry transformations of
eq. (\SuperSymmetry),
that this deformation preserves supersymmetry. 
The details of this are left for a future publication
[\CederwallLarsson], where a fuller account will be given.

\section\conclusions{Conclusions}We have reviewed and constructed
matrix theories describing situations where supermembranes end on
M-theory {\old5}-branes. Special emphasis has been put on
non-vanishing $C$-field, which is also where the new results are
found. 

There are some potential applications of the results, that will be
investigated in a future publication. One is to obtain the decoupling
from gravity in the limit of maximal $C$-field, the OM limit. For any
value of the $C$-field, we should be able to use our formulation to
derive the open membrane metric, which should arise naturally after certain
rescalings in the process of matching the truncated membrane
hamiltonian to the SYM one. 

One of our motivations for initiating this work was the prospect of
treating membrane boundary conditions in the presence of non-vanishing
$C$. In order for this to work, and to get information of the
generalisation of the string end-point non-commutativity to membrane
end-strings, one would need to find a generalisation of the approach
{\bf B} described above, so that the string nature of the boundary is
preserved. We have not been able to do this. An intriguing observation
is that there are two inequivalent cocycles extending an $su(N)$ loop
algebra to an affine algebra---the untwisted and the twisted one. The
zero-modes of the twisted affine algebra form an $so(N)$ algebra,
which certainly indicates a connection to approach {\bf B}. Further
investigations along this line of thought might provide interesting results.

\refout
\end